\documentclass[fleqn]{article}
\usepackage{graphics}
\begin{document}
\newcommand{\tab}{\hspace{5mm}}

\newcommand{\keywords}{quantum mechanics, Maxwell-Hertz electromagnetic theory,
detected signal} 

\newcommand{\PACS}{34.10.+x, 03.50.Kk, 84.47.+w, }

\title{Louis De Broglie's experiment}
\author{S.\ L.\ Vesely$^{1}$, A.\ A.\ Vesely} 

\newcommand{\address}
  {$^{1}$I.T.B., C.N.R., via Fratelli Cervi 93, I-20090 Segrate(MI)
   \\ \hspace*{0.5mm} Italy \\ 
   }

\newcommand{\email}{\tt sara.vesely@itb.cnr.it, vesely@tana.it} 

\maketitle

{\small
\noindent \address
\par
\par
\noindent email: \email
}
\par

{\small
\noindent{\bf Keywords:} \keywords \par
\par
\noindent{\bf PACS:} \PACS 
}

\begin{abstract}

Louis de Broglie's celebrated hypothesis transfers a problem 
of representation of optics to the quantum theories. Let us suppose 
that the fact that originated the problem in optics is the following, 
``The images obtained by optical instruments are limited by diffraction. 
Is the diffraction phenomenon exclusively restricting of the 
information contained in the image or does it have some physical 
meaning by itself?'' This considered as a problem of measurement 
of the radiative field does not pose the question of tracing 
well-defined processes in the sense of the light quantum concept. 
Despite this the application thereto of quantum formalism makes 
the essentially inferring statistical nature of measurement forecasting 
most important. We suggest avoiding the probabilistic conception 
of measurement by replacing the wave-corpuscle dualistic principle 
with the duality one. The latter is a basic feature of projective 
geometry. Referring to it we shall try to include the diffraction 
in a geometrical model of optics and interpret in that coherent 
schema the point -plane ``wave'' dual solution. We 
show what experimental substantiation the proposed duality has. 
Lastly, we again discuss the experimental basis of wave-corpuscle 
dualism.
\end{abstract}

\section{Introduction}

According to the literature \cite{messiah1999} in the 
thesis work of L. de Broglie \cite{debroglie}, 
'recherches sur la th\'{e}orie des quanta' the famous hypothesis 
that the same wave and corpuscle dualism as is present in light 
may also occur in matter is formulated. The quantum mechanics 
\cite{vanderwaerden} accepts the correctness of 
the terms because it places them there where they are expounded 
in the manuals. Although it discusses the consequences, it does 
not seem to assess the assumptions. Let us consider two aspects, 
one about the widespread use of calling it a hypothesis and the 
other about the undulatory and corpuscular representation of 
light.

Precisely it is written that de Broglie formulated a hypothesis. 
It concerns a metaphorical analogy of light with matter. We believe 
that, thus formulated, the hypothetical analogy renounces interpretation 
of the facts. We want to explain this. A mathematical hypothesis 
may be formulated within the framework of a unified theory of 
light and matter \cite{ditchburn1991}. A theory which 
is not mechanics in short because this does not explain typically 
luminous effects \textit{if there are any}. But in 1924 there is still 
no common theory of light and matter \cite{mehra2000}. 
Therefore at that time the hypothesis is not a conjecture formulated 
within a theory. It might be a supposition on which to base a 
new research program. As it concerns physics it should rest on 
a natural phenomenon. But the hypothesis precedes in time all 
the experiments aimed at verifying it. Indeed, on the one hand 
it does not grasp any evident resemblance between light and matter. 
For example, the sun reflected by the burning glass mirror \textit{does 
not} bombard ships exactly like a cannon ball. On the other hand 
the oft quoted scrupulous experiment of C. Davisson and L. Germer 
\cite{davisson1927} is of 1927. This experiment confirms 
the de Broglie hypothesis in the sense that based on electronic 
diffraction measurements from a nickel crystal it associates 
a wavelength with the collimated electronic emission. This could 
be used to confute the conclusions reached by J.J. Thomson about 
the corpuscular nature of cathode rays just as well as it was 
used to prove the undulatory nature of matter. In short, in 1924 
the hypothesis anticipates both the theory and the verification 
experiments.

The second imprecision of the literature concerning the contribution 
of de Broglie seems to us to consist of a reversal of the order 
in which the words appear. In our opinion the hypothesis would 
state more correctly, ``the same dualism of wave and corpuscle 
as is present in matter may also occur in light''. Indeed, 
not only does it not deal with modeling mechanics in accordance 
with a dualistic theory of radiation already formalized, but 
in addition both the emissive material conception of I. Newton 
and the elastic one of A. Fresnel belong to dynamics so that 
exactly the dualist concept of light is mechanical. But in 1924 
there are at least two formalisms foreign to the conventional 
mechanical schema, the electromagnetic theory and the quantized 
energetics of A. Einstein \cite{einstein1917}. The first 
theory concerns more the propagation of electrical phenomena 
than the characterization of ether; the second more the energy 
balance of elementary processes \cite{einstein1907} 
than corpuscularity. Since the principle of interference of Huygens-Fresnel 
can be fitted on electromagnetism \cite{towne1988} and 
in addition Einstein himself hastened to establish an equivalence 
between mass and energy \cite{einstein1905}, it may 
be thought to trace back the two attempts in the already broken-in 
tracks of mechanics \cite{debroglie1949}. This, which 
more than a hypothesis we would like to call an idea \cite{bohr1949},
 is valid in the absence of reasons justifying the 
attempts to develop more original light theories.

In this work, which is not historical despite the problems faced 
being so, we shall briefly confront dualism in physics. Then 
we will seek to explain how geometry allows justification of 
the dual aspect of light phenomena in our opinion. In literature 
other possible geometrical interpretations \cite{feoli2002} 
are discussed. Lastly, we will seek to give a physical meaning 
to ``de Broglie's hypothesis'' in accordance with \textit{the experimental 
results obtained earlier by E. Abbe} \cite{abbe1989}. 
This is the reason for the title we chose.

\section{Dualism as a contribution to unification}

De Broglie's contribution to physics is to have contemplated 
two different entities, light and matter, and to have conceived 
that wave-corpuscle dualism can unite them \cite{cushing2000}. 
This means that dualism is referred to undulatory or corpuscular 
pictorial descriptions but at the same time unifies the two entities 
light and matter.

In quantum mechanics the conflict between the wave and corpuscle 
concepts is resolved by postulating the principle of complementarity 
\cite{born1969}.

As to the essence of the two entities light and matter which 
the theory treats monistically, we see two possibilities. Either 
both exist in the outside world, which we as physicists admit, 
and dualism is accessible to the experiment. Or division of things 
in two (or more) classes naturally is not \textit{given} and dualism 
between light and matter is equivalent to that between wave and 
corpuscle. Let us take the first case. If we consider the distinction 
between light elements and matter elements a natural fact, in 
physics we should discuss the experimental consistency of this \textit{not 
unifiable} dualism between at least two types of distinct substances. 
Second case: but if we consider linking between experiences and 
pictures the work of physicists, dualism (or pluralism) takes 
place at the level of \textit{our} understanding of the experiments; \textit{we} 
are referring certain experiments to mechanics and we judge those 
remaining more linked to sensorial impressions. At this level, 
theories different from mechanics describe effectively different 
aspects of the external world. Since in addition dualism is not 
among the facts accessible to experimental investigation, neither 
does the level of its insertion among objects of the external 
world and their intellectual representation have an objective 
substantiation. In this case a dialectic comparison between alternative 
theories and conventions might not be contradictory.

To summarize, physics is the science of physicists but it is \textit{also} 
a science which explains phenomena occurring in nature \cite{matzkin2002}.
On de Broglie's thesis it can be substantiated how 
he does not propose at all to put light manifestations in relation 
with material bodies. Rather, in the 1924 work he presumes that 
theories are platonic ideas and formulates his unifying hypothesis 
for two preexisting theories. He chooses analytical mechanics 
and geometrical optics. Geometrical optics is distinguished from 
analytical mechanics as long as the methods of the latter are 
not applied to the former. And this ensures the success of unification 
even if it perhaps makes superfluous formal verification of dualism 
separately for the light theory and for the matter theory.

Although it is innovative to introduce dualism in a theory, the 
dualistic aspect considered is not specific to light theories. 
To clarify what we mean, let us consider the contrasting pictorial 
descriptions which would unify the dynamics with optics. They 
are termed pictorial because they evoke in us familiar concepts. 
Their conflict \cite{born1969} precedes the formulation 
of de Broglie's hypothesis because it expresses a contradiction 
between being and \textit{movement}. The dilemma between the above 
mentioned concepts is a pillar of western philosophy and was 
fed already in ancient Greek philosophy by two schools, one termed 
Eleatic and the other Heraclitean. For the purposes of this article 
we agree that the first of the two terms, being, specifies an 
actual (``realised'') property on condition that it be permanently 
preserved. The mathematical magnitudes introduced are used to 
represent a state. The other term, movement or ``becoming'', expresses 
a relationship between spatial positions of the same thing, for 
example along a trajectory, or indifferently between two or more 
things \cite{russell1981}. Mathematically if the relationships 
represented refer to pairs of facts they are termed binary, otherwise 
ternary et cetera. The variables represent not necessarily \textit{punctual} 
relative localizations. \textit{The state of things cannot be part 
of the representation}.

In classical mechanics, with being are associated the causes 
of motion represented mathematically by means of vectors or tensors 
(state of motion) while with becoming are associated displacements 
in geometrical space (kinematics). The data available allow establishing 
for each specific application which is the best adapted representation 
and therefore contradictions do not arise. To mix the two pictures 
it is necessary to abolish the conventional distinction between 
applied vectors (forces) and free vectors (displacements) which 
is what takes place by graphically representing both on the same 
sheet of paper.

Contrary to mechanics optics, as a theory, is contradictory. 
The great merit of the radiation theories of Einstein and Maxwell 
lies just in \textit{avoiding} contradictoriness by directly applying 
a mathematics to a preselected class of light phenomena. In this 
work we shall consider the theory of the second because \textit{in 
the present state of telecommunications development} it is applied 
to the signal. Einstein's theory of radiation, since it puts 
on an equal footing modulation and statistical fluctuations, 
cannot include images like those of television when the television 
is tuned to a channel.

\section{Contradictory pictures in optics}

Since from the beginnings of logic the choice of basing a hypothetical-deductive 
theory on a dualistic principle appeared debatable, in the previous 
chapter we identified in pluralism a possible alternative to 
a well-founded theory. Optics is not a theory but a discipline 
which gathers together a large number of unrelated facts. The 
known facts, of which it is possible to specify the observation 
conditions, are truly many because a goodly part of our information 
about the external world is linked to the visual channel. The 
number of explanations given is also considerable. If we remain 
with the descriptions having objective corroboration, the first 
difficulty lies in disentangling the physiological from the physical 
contribution to visual impressions. The theories of colors are 
emblematic.

\subsection{Graphic optics when diffraction is a small effect}

\tab If we go back to the history of optics, as does de Broglie in 
the thesis, we can observe that in the same period the two representations 
he calls ``undulatory'' (from C. Huygens's; without periodicity principle) 
and ``dynamic'' (from Newton's theory) are strictly relevant to 
the \textit{same} experimental base and the opposition of ethereal 
wave-trains and luminous corpuscular matter takes place \textit{only} 
at the interpretive level \cite{mach1982}.
The physical content of optics commonly termed geometrical is 
summarized in graphic memoranda of the conditions of operation 
of well diaphragmed instruments. These are used to view sharp 
images of known objects and are first of all the pinhole camera, 
then lenses, mirrors, telescopes and the like. The majority of 
the graphic memoranda analyze their operation in terms of reflection 
and refraction with recourse to the radial geometric element. 
Mathematically speaking the theories developed on rays are linear. 
In them the so-called optical aberrations are considered a small 
error which is allowed for by making corrections to the linear 
theory. If by applying differential calculus the graphic trace 
is corrected to higher orders, smooth curved lines replace the 
rays.

In the same period as Newton, Huygens was able to trace the same 
stigmatic images more elaborately by using a surface \textit{element} 
associated with the ray-optical path, instead of the ray itself. 
He showed us that by assuming fronts with some extension instead 
of single trajectories it is possible to justify graphically 
even two distinct stigmatic images if the fronts have two different 
envelopes. Strictly speaking we could not call stigmatic a split 
image and would judge unusable for the purposes of observation 
an optical instrument which splits the objects observed. But 
Iceland spar appropriately cut and finished optically shows better 
than other crystals that it can form a neat double image. Let 
us describe the fact. Resting the crystal on a sheet on which 
a point is marked, the image thereof appears double and, rotating 
the spar, the duplicate executes one rotation around the ordinarily 
refracted point, on a visibly higher plane.

Huygens had believed that the split image was not explainable 
otherwise than by also assuming the propagation of non-spherical 
wave fronts inside the spar. Newton objected that it was a matter 
of understanding if one should assign the responsible property 
of the splitting of the image to the light matter or to an ethereal 
fluid. In the second case he didn't understand why ether should 
possess some anisotropy inside the spar and not show any in intergalactic 
space. Actually the polarization could be understood in Newtonian 
terms and it was thus that E. Malus fixed it for posterity.

Again according to history, Newtonian optics did not reveal itself 
insufficient to explain the birefringence mentioned above but 
failed on the diffraction experiments of Fresnel.

Diffraction as a phenomenon had already been accurately described 
by Newton. But no one believed that, when seen under high magnification, 
it might prove to be a remarkable effect. In addition, the supposed 
inadequacy of geometrical optics to account for Fresnel's experiments 
has been used more for setting precise limits of validity of 
the graphic analysis handed down than for unequivocally identifying 
the fact to which to attribute the fanlike opening of a diaphragmed 
light beam. Dispersion, interference, diffraction, diffusion\dots  
all contribute to blurring the image. Although, contrary to Newton, 
we hesitate to assign each as a separate property to light, nevertheless, 
if we want to allow for the opening of the beam like a fan we 
must choose to deal with one of these phenomena.

We are inclined to choose diffraction under the conditions of 
J. Fraunhofer, first of all because the experimental conditions 
are well defined and then because this diffraction can be treated 
mathematically. Ours is a choice and as such is arbitrary. R.W. 
Hamilton made another choice. He found a mathematical relation 
between a normal congruence of rays and a family of surfaces 
\cite{klein1973b}. He also showed that integration of the 
differential equations of dynamics is in relation with the integration 
of a first order partial differential equation (the one for the 
generating function). The above relation can receive any convenient 
physical interpretation \cite{bailey2002} provided it 
is remembered that the phase space and that of graphical optics 
are not isomorphous \cite{pauli1973}. Hamilton applied 
the mathematics developed by himself to the conical refraction 
of biaxial crystals.

Conical refraction is so called because the image of a point 
observed through a biaxial crystal, instead of appearing split 
appears widened like the base of a cone. Fraunhofer's diffraction 
takes place under quite different conditions, i.e. even in the 
absence of an interposed dielectric material. For this reason 
we speak of a choice.

\section{The problem of what moves in optics}

Until now we have said that behind the early theories of light 
there are \textit{instrumentation} manufacturing technologies which 
pursue magnification on condition that images shall be similar 
to the original in form and colors. Even then opticians were 
pragmatic, they assumed that the graphic prescription which allows 
analysis of the image more conveniently was the most natural 
in a teleological sense. Since Newton left us a (reflector) telescope 
they considered it certain that he had understood the nature 
of light better than Huygens.

As concerns the Newtonian interpretation of the rays as particle 
trajectories it is perfectly useless. It is not useful for improving 
the practice of processing optical glasses of refractory telescopes. 
It is not even useful for further characterizing by experiments 
the material nature of light; the proof of this lies in the convincing 
explanations of diffraction based on the periodicity principle 
due to T. Young and Fresnel.

Maxwell' starting point is not oriented toward improvement of 
instrumentation but, in modern terms, towards coherent representation 
of the signal received when we can disregard instrumentation 
particularly. Not in judge's clothing but in a lawyer's, we insert 
a digression to give an idea of the theoretical conception. Maxwell, 
contrary to Einstein, considers only the phenomenology linked 
to movement. Consequently he does not characterize and does not 
interpret the moving object which we are calling a signal (to 
distinguish it from a tangible body). A signal can be received 
through being propagated along an electrical or optical transmission 
line. But if it is irradiated in space it can be detected mechanically, 
chemically, electrically, magnetically, thermally or optically. 
Since characterization of the signal is not the purpose of the 
theory, the latter is the more general the more extended is the 
class of signals considered. This generality cannot be unraveled 
as long as the specific job attributed to optics is \textit{to explain 
the transition of the object to the image}. But this transition, 
once separated from the process of production and detection of 
the signal, is a sui generis movement. To the touch it is established 
that an object standing in the laboratory and visible to the 
naked eye does not move due to the fact that it is observed directly 
rather than with the aid of an optical instrument. And yet the 
image of the object in toto is formed visibly elsewhere. Either 
microscopic particles transit from the position of the object 
to that of the image, a case in which mechanics can be applied 
as is to light (theory of F. Hasenoehrl \cite{hasenoehrl1904}) 
or the light can involve matter more ineffably. In the second 
case for the sake of argument the movement has no equivalent 
in classical kinematics.

Now let us go back to Maxwell. He conceives interference as proof 
of the \textit{immateriality} of light. He reasons more or less thus: 
the description of fringes, requiring evaluation locally of the 
algebraic sum of two quantities -- those which interfere -- testifies 
against the emissive theories, at least as long as mass is the 
substantial characteristic. Indeed, usually to mass is attributed 
an essentially positive magnitude so that, contrarily to the 
electric charge, there is no analogy of Faraday's cage capable 
of screening it. Following M. Faraday further, he lets himself 
be guided by spatial considerations; phenomenology develops entirely 
in our world and therefore describing it has sense at first. 
At that time, this might have seemed a foregone choice \cite{manning1977}
 \cite{rucker1977}.

A mechanical illustration is also part of Maxwell's original 
interpretation. To make plausible the distribution of light intensity 
on the immaterial fringes, he identifies electromagnetic ether 
with luminiferous ether. The latter is a hypothesis on the indirect 
role of a medium during transmission; ether is a medium which 
is subject to elastic vibrations while light traverses it and 
therefore transmits the dynamic energy of light along definite 
directions. To \textit{dynamically} \textit{characterize} the propagation 
of light it is necessary and sufficient to specify the state 
of motion of ether in accordance with the theory of elasticity. 
Of course the dynamic characterization could continue to reveal 
itself perfectly useless in applications.

After Maxwell it was noticed that, whatever ether is, contradictory 
material properties should be attributed to it. Therefore it 
was preferred not to consider it a real entity. It follows that 
vacuum \textit{is not an entity} replacing ether but is a \textit{non-entity}. 
Denying the existence of electromagnetic ether and replacing 
it with nothing, nothing takes on a state of motion. In other 
words, an elastic vibration energy associated with light propagation 
causing the appearance of the fringes is lacking. Nor does thermal 
conduction account for the light distribution observed experimentally 
in a manner which might be consistent with electromagnetism in 
a vacuum. Indeed, the thermodynamic interpretation of the flux 
of light energy assumes that the thermodynamic state variables 
are functions of the spatial coordinates and possibly of time. 
It takes on this dependence to handle the diffusion of heat by 
conduction from the hotter zones to the cooler zones of a body. 
But the coordinates describe the \textit{thermal state} of the conductor 
\cite{maxwell1954}.
Part of the above discussion aims at concepyually identifying what 
might move while conveying energy in the absence of an interposed 
medium. In electromagnetism every general solution (without 
boundary conditions) represents a movement of permanent type. 
Although these movements have been interpreted mechanically as 
waves of ether, they do not have a mechanical characterization. 
For example, they have nothing in common with surface waves like 
those of the sea nor with wave fronts like those generated in 
a pan by emersion of a cork nor with the vibrations of tight 
ropes. They represent a movement dissociated from any other characterization. 
Maxwell explains this in the Theory of Heat, especially in chapters 
XV and XVI, first paragraph.

The crucial point is that \textit{the equations remain valid regardless} 
of the outcome of the discussions on the essence of light because 
the vacuum hypothesis is perfectly compatible with the geometrical 
representation of movement. By geometrical movement is meant 
in this case the transformations of space in itself \cite{klein1987}.

\subsection{Louis De Broglie's phase waves are not electromagnetic}

Let us recall that Maxwell's equations \textit{represent} movements 
while the interferential fringes are the \textit{fact}, i.e. the ``thing 
which interferes''. Experimentally, the fringes are said to be 
of equal thickness if they appear localized on the surface of 
iridescent bodies; in this case light intensity can be put in 
direct relation with the thickness of material on which it strikes. 
But if they are observed under Fraunhofer's conditions they are 
said to be focused at infinity, meaning that they are not at 
all localized because there is no place localized at infinity 
where their source lies; consequently something is lacking to 
the finite in relation to light distribution. This fact can be 
interpreted in a manner compatible with electromagnetism as a 
property of light or as a property of the object-optical instrument 
pair or otherwise just because Maxwell didn't write a theory 
of the fact, that is he didn't represent interference fringes. 
But the \textit{interference fringes} \textit{immateriality} requirement remains 
a basic requisite because it brought about Maxwell's initial 
choice of representing not states but movements \cite{born1925}.
 To keep in the empty space the hypothesis that 
these fringes are not substance is equivalent to denying that 
optical instruments function like cannons. More precisely, it 
is denied that something is projected into the image like a cannon 
ball is projected from the initial position to the expected final 
one -- provided aim is good.

In the thesis de Broglie does not hypothesize that optical images 
can behave like shot cannon balls because this is contrary to \textit{all} 
experimental evidence. Rather he hypothesizes that an energy 
transfer takes place because of a convective motion of emitted 
photons, motion which admits a statistical mechanical model
\cite{boltzmann1999}
but possibly without satisfying the principle of 
equipartition of energy. With the statistical mixture of photons 
he then dualistically associates phase waves such that the instruments 
focus \textit{with scattering}, i.e. with some blur. These phase waves 
represent the intensity patterns which the diffusion of photons 
alone doesn't justify. The interferential fringes observed experimentally 
are no longer a fundamental hypothesis of the theory but are 
the ``visible result'' of a random reality to which the theory 
accedes. In other words a phenomenon (i.e. the interferential 
fringes) is not observed but the statistical frequency with which 
something absolutely inaccessible to the experiment occurs.

In short: the guiding waves, of which de Broglie speaks, are 
not equipollent to the electromagnetic ones because the former 
justify interference with scattering of photons while the latter \textit{do 
not at all explain} interference; Maxwell limited himself to ascertaining 
this \cite{maxwell2001}. Sometimes he calls it interference 
and sometimes more generically diffraction.

\section{Periodicity as a property of light}

The intricate subject of the essence of light was not faced by 
de Broglie in the thesis as we remarked above \cite{hund1984}.
For this reason we shall limit ourselves to a few considerations 
on its periodicity property. In mechanics when it is a matter 
of phenomena accessible to the experiment, i.e. reproducible 
in the laboratory, it is possible to verify their degree of periodicity 
among other ways by comparison with a periodic device having 
a period near the one investigated. If the periodicity of the 
phenomenon to be tested is not simple, the intent can be laborious. 
If in the long run we really can't distinguish any periodicity 
in the phenomenon it is usually agreed that it is random. Then, 
as no evolution law can be given because of the endless becoming 
of the phenomenon, we have recourse to statistics.

In optics the interferential fringes as such result indeed from 
the comparison between two radiations as much as possible alike. 
This does not effectively mean that we possess a periodic reference 
device but that the source is the same. The experiment allows 
attributing neither periodicity nor fortuitousness to the phenomenon. 
Let us explain this point. A spatial tessellation with a Platonic 
solid which is a covering can be represented geometrically by 
means of a finite group of movements of space in itself. Each 
of these movements can be expressed by parametrizing an appropriate 
function of the spatial variables. These functions, written in 
terms of the real parameter, are commonly called periodic. Just 
to interpret the real parameter as time, the geometrical regular 
arrangement can be transferred to temporal evolution. The space-time 
correspondence is mathematically fixed but there is a dissymmetry 
between the spatial and temporal variables. Time remains the 
parameter of a spatial transformation. It is not deduced that 
the phenomenon described is periodical in the sense given to 
the word in mechanics. Therefore this description does not make 
physical hypotheses regarding the nature of the fringes. On the 
contrary, a theoretical formulation assigning a period to a mathematical 
function of random variables does not specify the nature of the 
phenomenon but is contradictory.

\subsection{James Clerk Maxwell doesn't attribute basic colors to 
light}

Among those who experimented and reflected on optics there is 
notoriously Newton \cite{newton1979}. He attributed the colors 
of the rainbow to light and compared them with the diatonic musical 
scale, tracing back sounds and light to states of motion. His reasoning 
is subtle; here we can quote a definition as an example: ``The 
homogeneal Light and Rays which appear red, or rather make Objects 
appear so, I call Rubrifick or Red-making; those which make Objects 
appear yellow, green, blue, and violet, I call Yellow-making, 
Green-making, Blue-making, Violet-making, and so for the rest. 
And if at any time I speak of Light and Rays as coloured or endued 
with Colours, I would be understood to speak not philosophically 
and properly, but grossly, and accordingly to such Conceptions 
as vulgar Pepople in seiing all these Experiments would be apt 
to frame. For the Rays to speak properly are not coloured. In 
them there is nothing else than a certain Power and Disposition 
to stir up a Sensation of this or that Colour. For as Sound in 
a Bell or musical String, or other sounding Body, is nothing 
but a trembling Motion, and in the Air nothing but that Motion 
propagated from the Object, and in the Sensorium 'tis a Sense 
of that Motion under the Form of Sound; so colours in the Object 
are nothing but a Disposition to reflect this or that sort of 
Rays more copiously than the rest; in the Rays they are nothing 
but their Dispositions to propagate this or that Motion into 
the Sensorium, and in the Sensorium they are Sensations of those 
Motions under the Forms of Colours.''

Hereafter we shall prefer to refer to colors rather than frequencies. 
Newton concluded that sunlight is composed of single spectral 
colors from the fact that it is dispersed by a glass prism in 
the colors of the rainbow. Since glass technology hadn't the 
interest it later acquired, he ignored that dispersion depends 
on the chemical composition of the glass.

Maxwell on the other hand was in a position to observe that, 
if that which has the property of color is light but there are 
pigments which radiate only light with well-defined coloring, 
then matter also has the property of color. In the case of rarefied 
gases excited electrically, color radiates homogeneously and 
isotropically trough space, i.e. patterns are not formed. If 
we manage to obtain persistent light, which does not happen by 
subjecting a rarefied gas to any stresses, color is also largely 
independent of external circumstances. It can be asked if this 
characteristic allows using the phenomenon to define a physical 
time relatively to duration or periodicity. The periodicity introduced 
here is not in relation with that of the last paragraph; rather 
it should be traced back to the stability of the feeding device.

It seems that Maxwell was not entirely a stranger to the problems 
linked to flow of time, timekeepers and movement. In the last 
pages of the Theory of Heat he asks if, admitted that matter 
consists of molecules, the latter can be harmonised with the 
``light unison''. For sure there he doesn't link resonance to 
a time arrow. However that may be, in electromagnetism to assume 
the existence of fundamental components of a particular frequency 
is the equivalent of raising the method of resolution by expansion 
into autofunctions to a principle of physics.

On the contrary, if a physical meaning is not attributed to said 
mathematical method, it can be replaced to all effects with the 
integral transform method \cite{Maxwell1954b}.

\subsection{Electromagnetism is not a spectral theory}

``Electromagnetic waves'' thought of as propagating ``monochromatic'' 
solutions of the wave equation exist subject to foundation of 
a \textit{spectral theory}. As long as in electromagnetism movement 
is attributed to ether, dynamics supplies the laws of motion 
of ether and perhaps explains the color of light. If movement 
is geometrical the only laws, those of transformation, are supplied 
by geometry.

A spectral principle for explaining interference and diffraction 
comes into conflict with Euclidean geometry adopted in particular 
by Maxwell. Indeed, in assuming the existence of basic physical 
components it is imagined that they appear progressively or suddenly 
with increase in resolution. If the principle is effective it 
attributes absolute magnitudes to the basic constituents. But 
Euclidean geometry does not allow introducing absolute magnitudes. 
To explain, if by using a telescope with better image definition 
we can no longer distinguish canals on Mars, the conclusion that 
these were artifacts at the previous degree of resolution is 
in agreement with the Euclidean principle of similitude. The 
alternative of considering the canals as real structures visible 
only by special enlargements is in contrast with the above mentioned 
principle. In this sense the geometrical theories unified with 
the spectral ones are contradictory.

Below in this work we shall no longer attribute several spectral 
components to light. But this choice involves only the theory 
since all optical instruments are dispersive. It must only be 
remembered that some disperse very little, and others -- interferometers 
and spectral analyzers -- disperse more.

\subsection{Interference in optical geometry}

Traditionally, taking on the Huygens-Fresnel principle, optics 
manuals inaugurate a new chapter dedicated to \textit{physical or 
undulatory optics}. But this formulation must not be misunderstood. 
No manual intends to answer the question whether it is necessary 
to introduce this new basic principle to explain the phenomenology 
of the new chapter. To judge whether geometry suffices or whether 
the observations require development of a spectral theory is 
the concern of one proposing to write \textit{one} coherent theory. 
Observations on which one might rest the first hypothesis are 
not lacking. For example, in 'On the Estimation of Aperture in 
the Microscope' E. Abbe writes, ``Let us assume that the fineness 
of a striation, a grating or a diatom is at the limit of instrumental 
resolution for \ensuremath{[}a microscope of\ensuremath{]} a given aperture 
\ensuremath{[}\dots \ensuremath{]}, then always and only two diffracted rays 
contribute to the image \ensuremath{[}observed with the eyepiece\ensuremath{]}. 
In clear field this would be the direct ray, and one spectrally 
decomposed, \ensuremath{[}the first is the zero order and the ``colored 
one'' is the first order of diffraction\ensuremath{]}. In this case 
the striation seems to be in clear and dark bands \textit{of equal 
width}. But we know that with \ensuremath{[}an objective of\ensuremath{]} greater 
aperture we would obtain a more complete image. In this the ratio 
between the widths of clear and dark bands would be much different.'' 
The italics are ours. In this passage Abbe explains that interference 
is an image and not generically a pattern; in addition that it 
is the image of \textit{any flat grating}, provided the instrumental 
resolution is insufficient. The result is that interference seems 
to be a property of light because it does not characterize the 
object observed although it is it's image. But the microscope 
observations serve to elucidate the structure of the bodies observed 
and therefore it is the bodies which possess the property of 
being visible.

\section{Ernst Abbe's hypothesis on what moves}

We saw that, according to Abbe, the interference pattern is an 
image. We repeat that the immaterial or ethereal image does not 
have existence independent of the thing (it goes without saying 
that, if the thing is not alight, the image depends \textit{also} on 
illumination). But microscopic observation is an observation 
of light only. More exactly, it is about the kind of light which 
in telecommunications takes the name ``signal''.

\begin{figure}
  \resizebox{8cm}{8cm}{\includegraphics{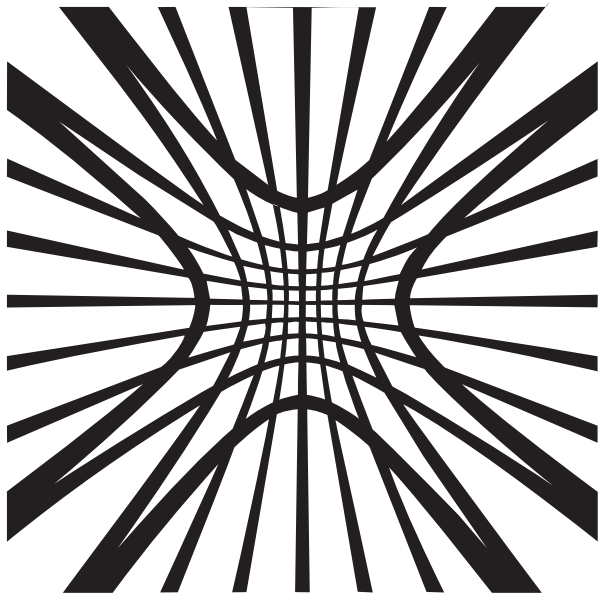}}
  \caption{Aplanatic distortion of a square grating.
Two arrays of orthogonal equidistant parallels are transformed into two arrays of homocentrical hyperbolas.  Aplanatic distortion may be observed looking through wide-angle lenses, in case there are no convergency errors.
    }
  \label{fig_1}
\end{figure}

Now let us say what moves according to Abbe. If there is no other 
way to accede to something but to observe it with the aid of 
an optical instrument, it is illusory to assume that the image 
has the visual appearance which the thing would have if we could 
observe it directly. This justifies the aptitude to also consider 
images those figures which certainly do not reproduce \textit{in 
all} detail the structure examined. With this qualification Abbe's 
aptitude consists of assuming that \textit{the image is the linear 
transform of the anti-image}. This linear transform limited to 
the signal represents the displacement of competence of optics. 
Analysis of the displacement totally disregards all information 
conveyed by the signal, hence any question concerning its interpretation. 
In addition Abbe experimented with when the image is formed and 
how it depends on the lens numerical aperture; then he appraised 
within what limits one may speak of geometrical transformation 
and in this case how to express mathematically the condition 
of formation of the image. What he called ``condition of aplanatism'' 
is clearly an invariant of the transformation. One of his works 
on the subject is entitled, '\"{u}ber die Bedingungen des Aplanatismus 
der Linsensysteme' \cite{abbe1989}. In the figure \ref{fig_1}
we reproduce from that work the plate in which he calculates 
the aplanatic distortion of a square grating. Observing this 
plate through the lens of the microscope he saw a grating with 
square meshes. Abbe also explains in what this distortion differs 
from spherical aberration. In effect the aberrations which meet 
aplanatism are practically those of Seidel.

\section{Geometry as linear model of optics}

Geometric modeling assumes for its applicability that relations 
between facts are linear. Now, saying that a cannon and a mirror 
don't have the same use, we don't bring any evidence that the 
phenomenology associated with light manifestations of matter 
is easier to handle mathematically than that associated with 
combustion, friction or viscoelastic deformations. On the contrary, 
we foresee excluding from the model power applications such as 
the surgical laser and microwave ovens.

Among the reasons for which the cannon and the mirror don't behave 
in the same manner there is thus also the different use. The 
cannon is used for offensive purposes while the mirror is used 
as a detector of a signal-- the one sent by the object. In signal 
approximation the properties required to form reflected (or refracted) 
images can be attributed to light just because for the purposes 
of the \textit{interpretation} of the image the detector function 
can be disregarded. The image itself depends on the detection 
technique used (not only on the object which transmits it and 
the transmission channel) because it is possible to exploit a 
good number of different effects for detection purposes. Despite 
this we can take into account some instrumental characteristics 
in signal approximation. In this case we say that the receiver, 
including the detector, functions as a black box.

Below we seek to specify possible uses for geometric modeling 
in a theory of light in linear approximation. Geometry can take 
on one or the other of two functions. It can serve as a \textit{linear 
model of a relationship}. It will be our responsibility to choose 
this in such a manner that the model will be linear. But this 
does not imply that the relationship itself is analytically a 
straight line in Cartesian coordinates. Otherwise we can choose 
to use \textit{a geometry to describe our world}. Since our world 
is a geoid, if we intend that we represent it linearly by mapping 
a sphere on the plane, it will be well to clarify the meaning 
of linearity.

In the first case the geometric model cannot represent any peculiarity 
of our world. We avoid confusion if, instead of introducing space 
as an ineluctable modality of our perception, we immediately 
introduce geometry as a mathematics. In this case it is clear 
that the space represented supplies on the outer world the same 
kind of information which the cardinal number 3 supplies on the 
three animals Bianchina, Fiorella and Estella.

If, on the contrary we intend to accede to a physical aspect 
regarding the extension of the world and the access is \textit{experimental}, 
for example an optical detecting and ranging, again to avoid 
confusion, it is better to specify that we are concerned with \textit{topography}. 
Incidentally, a more traditional measurement than remote sensing 
is to lay a graduated ruler on the object while being careful 
to use a good surface plate common to the ruler and the object. 
Even in this case the graphic representation is conventional 
except that whoever uses it must succeed in orienting himself 
in the region represented by recognizing the elements on the 
map.

Abbe made the first choice. Indeed, he said that microscope images 
are not \textit{recognized} but \textit{known}. Then the linearity hypothesized 
for the model allows using the theory to represent the image 
of something. But the basic elements used to represent it are 
basic only geometrically. Linearity allows making use of the 
theory to interpret the image represented but it cannot consider 
it a linear approximation of the world we live in because there 
is no way to compare the image with the object. We make the first 
choice even when we propose a geometry as a diagram and a movement 
as a relationship. This is Maxwell's choice when vacuum is empty.

\subsection{A proposal of Felix Klein}

In 1901 F. Klein (No. LXXII in vol. II of the collection of scientific 
works; the article is quoted by M. Born and E. Wolf in ``Principles 
of Optics'', ed. 1991) studied the geometrical conditions 
under which the image appears similar to the original neglecting 
higher order differential corrections. This is the range of validity 
of the theory of Hamilton. He concluded that the theory is inapplicable 
to both the \textit{microscope} and the \textit{telescope}. He believes 
that the limitation on the applicability of geometry might be 
overcome by using projective instead of affine transformations; 
but he warns that projective optics would give rise to a theory 
completely different from Hamilton's. He says, ``The impression
that both \ensuremath{[}the geometries\ensuremath{]} allow reasoning connected 
to each other is recalled only by the marginal circumstance that 
linear relation appear at the end''. This author does not put 
forward projective geometry as a new typography but has in mind 
an application to physics of the mathematics which he himself 
developed.

Now we shall say what we want to show. At least one geometry, 
the projective one, doesn't require specifying whether the geometrical 
element chosen is a point or a plane. The reason is that all 
the projective theorems are dualizable by substituting the element 
``plane'' for ``point'' and vice versa. 
Incidentally, to speak of projective geometry as a geometry like 
Euclidean geometry or one of the infinite non-Euclidean geometries 
, i.e. with fixed curvature radius, is not quite correct. Indeed, 
projective geometry is not categorical, it is not just one geometry 
but many geometries, in fact infinitely many \cite{coxeter1989}.
Since the geometry chosen does not distinguish 
the geometrical point element from the geometrical plane element 
it is necessary that not even the physical interpretation make 
such a distinction. We can agree that the point represents the 
object, let's say in the specific sense of being the stigmatic 
image of the point, and the plane represents the associated wave, 
let us say in the sense of being Fraunhofer's diffraction pattern 
of the point. The above convention is equivalent to that for 
which the same uniform illumination characterizes both a disc 
and a luminous sphere; under certain conditions it is experimentally 
verified and under others it is not. We discuss this in greater 
detail below. For the time being we observe that this interpretation 
of wave-corpuscle dualism as a dual solution can be faced with 
the original experiments of Abbe and his interpretation of them. 
In addition, since said geometry admits the biratio as a basic 
invariant, it is not possible to introduce basic physical components. 
Abbe himself in 'On the estimation of the aperture of the microscope' 
takes the numerical aperture from the condition of aplanatism 
and shows that the expression of aplanatism is derived from a 
biratio. The same expression of Abbe can be found regardless 
of any consideration on the microscope in projective terms. This 
is done for example on the reference text indicated by Klein 
for projective geometry, 'Die Geometrie der Lage' by T. Reye 
\cite{reye1898}. Then the projective transformations 
of space in itself allow a mathematical representation alternative 
to the one proposed by Hamilton in which relative movements or 
displacements in our world are associated with transformations. 
This representation of movement contrasts with that proposed 
by Maxwell only as concerns the possibility of connecting the 
electrical quantities with the mechanical units of measure. This 
will be fully clarified below in this work for the unit of length.

\section{Images and their movements}

According to Klein, the most general movement is represented 
considering the group of projective transformations of space 
in itself according to the Erlangen program \cite{klein1973a}. 
In this work we shall not apply projective transformations of 
space in itself. Rather, we shall first describe the facts and 
then indicate which relationship to represent by means of mathematics.

We refer below to catoptrical systems for definiteness. The reason 
is that the specular surface of a mirror functions only as a 
screen and reveals (usually) a single image of the object, which 
is the one reflected.

\subsection{Localized images}

First let us consider a flat mirror. If it is in a room of a 
dwelling and we place ourselves in front of it we will not identify 
light rays but will see ourselves reflected in the mirror and, 
behind our image, a room of the same type as that which we know 
to exist behind us. Truly, since our world is a bit asymmetrical, 
in the world which appears in front of us the right is taken 
for the left. Although there is no clear way for believing that 
the world of the mirror is other than ours, someone might believe 
it different because inaccessible. The ``other world'' seems separated 
from ours by the impassable surface of the mirror only because 
the image is virtual. But any real image is physically accessible 
to us. Let us clarify further; the real image does not exist 
without the object and the instrument capable of detecting it 
but when it exists it is topographically locatable.

Let us discuss accessibility better. If we place ourselves before 
a concave spherical mirror, let's say with a diameter of about 
twenty centimeters, equipped with a ball of foamed polystyrene 
and a pencil, we observe the \textit{extension} of the image of the 
ball by going through it physically with the pencil. The image 
has extension in our world but we run through it without encountering 
obstacles because the mirror does not materialize it. Impenetrability 
can characterize bodies but not images. The same experience can 
be repeated on the real image refracted by a converging lens. 
But it is necessary to use the dioptric surface as a support 
for the image. Then the image visibly occupies a position with 
respect to the dioptric surface and the eyes settle automatically 
on the details observed. We know that an image can be detected 
by placing a flat screen of opaque or translucent material approximately 
in that position; the image then appears flat and more or less 
sharp. The second procedure is usual using dioptric systems but 
unacceptable for mirrors. For the same reason light diffused 
by objects cannot be diaphragmed so as to be ``paraxial'' using 
catoptrics. Clearly, every time the image is real it diffuses 
the light which illuminates the object for all its extension 
as though it were the surface of the object itself.

If we want to generalize this observation by hypothesizing that 
even virtual images are in our world, we cannot use Euclidean 
geometry because this geometry does not allow representing inaccessible 
regions.

Now we want to make a distinction which seems important to us. 
That real images are formed in the world in which we are is a \textit{fact}. 
Adopting a geometry such as to describe in the same way all locatable 
images is \textit{reasonable}. Detecting the movement of a body is 
on the other hand \textit{a measurement operation}. Even if we provided 
for representing the path with the same geometry used for representing 
images, they are facts, this is and remains a useful concept. 
In other words neither the trajectory nor the instantaneous speed 
with which a point travels over it are optical images. This is 
true apart from the priority we like to establish between concepts 
and things.

\subsection{Movements of localizable images}

In this paragraph we are not truly concerned with the optical 
detection of moving objects, rather our purpose is confined to 
examining one effect of increasing distance. As an example we 
can follow the movement of the point of the pencil looking in 
the concave mirror mentioned above, and supposed in relation 
to a graduation integral therewith. Qualitatively the correspondence 
between movement of the point and of its image is the following. 
While we cover the distance between us an the focus of the mirror, 
we see the image of the point move from the initial position 
in front of the mirror toward us until we no longer see it clearly 
because it is so close. Then we lose sight of it. Imaginarily 
it moves behind us more and more. When we pass through the focal 
point with the point and proceed towards the vertex it may happen 
that the image of the point enters into the visual field from 
the edge of the mirror as though it had traveled an elliptical 
trajectory; if on the other hand the alignment is good the contour 
of the stigmatic image forms visibly back-to-front behind the 
mirror. The image failed beyond the surface of the mirror and 
is separated from us by the specular surface. This surface, through 
which the point and its image finally appear to touch, is the 
only physical impediment to continuing of the movement of the 
pencil point. If the mirror is spherical, in addition to the 
one on the surface there is another point of our world in which 
object and image touch. It is metrically characteristic of the 
mirror and is quite localizable. If the mirror is parabolic, 
there is a single point where object and image touch et cetera.

If we had studied the path of withdrawal instead of the path 
of approach, for example on the ball, having it moved by someone 
behind us but in such a way as to continue to receive its image, 
we would have seen it proceed ever ``more slowly'' toward the 
focus as a limit point and flatten increasingly. All this is 
applied to the movement of the image without differing conceptually 
from the kinematics of fluids. As the only device it is necessary 
that each observer hold immobile his point of view of the mirror. 
A repositioning of the mirror in relation to the observer implies 
indeed a change of reference, the graduated scale being supposedly 
integral with the mirror.

\subsection{The information contained in the stigmatic image is that 
contained in the wavy image}

This short paragraph doesn't concern so much the image in 
geometrical approximation as it does analysis of the information 
contained. Up to this point we have introduced the stigmatic 
image of the nearby object not diaphragmed in diffuse light. 
The image is assumed to be reconstructed point by point, one 
plane at a time from the object's shell. Although usually in 
optics focal rays are represented, geometrically the transformation 
is a central projection \cite{pohl1976}. For this reason 
optics texts call ``collineation'' the image constructed according 
to the principles of geometrical optics. The stigmatic image, 
i.e. the one reconstructed in linear approximation, is not flat 
if particular conditions such as those relative to focusing on 
photographic film are excepted. We showed a way of ascertaining 
whether the specular image also reproduces the spatial extension.
Whoever has read about holography will not be surprised \cite{hecht1998}.
Among the more common holograms are small reflecting 
plates which, when uniformly lighted, appear as photographs having 
parallax. This moving effect of the planes according to their 
depth is codified interferometrically during recording of the 
hologram. The procedure consists essentially of impressing on 
a film the pattern of interference between a reference radiation 
and the diffuse light from the holographed object. A complicated 
pattern remains impressed on the film. After development the 
plate is capable of regenerating the ``wave'' diffused 
by the object by modulating a striking radiation of the same 
type as the one used as reference during recording. It appears 
that in the approximation in which we suppose the pattern etched 
on the perfect plate, or the mirror free of aberrations, we can 
explain the \textit{not flat} surface of the image by making use indifferently 
of the properties of radial propagation assumed for the light, 
or of the wavy characteristic attributed to radiation. The pictures 
are complementary but the decodable information from the signal 
received is \textit{the same}. From the analysis made it would seem 
that we could really infer that a dual nature could be ascribed 
to light in the sense of de Broglie. But 
this is not so. This dual nature must be attributed to the receiver 
paired with the respective coder in the sense of the principle of complementarity.

Setting aside the theories 
of light, the signal received by the mirror can also be interpreted 
as though the instrument performed a central projection.

\subsection{The delocalized image}

In Abbe's analysis the image of the light point or the uniformly 
lit plane is the extreme case \cite{ditchburn1991}. 
In the example of the mirror we begin to make plausible that 
geometry is sufficiently descriptive of what is observed. Let 
us assume that the object, which now need not be symmetrical, 
and rather it is better that it not be, while it continues to 
be illuminated in a diffuse light withdraws behind us to the 
horizon while we, integral with the mirror, continue to receive 
its real image. In practice, whatever the form of the object, 
we shall finish by seeing a small luminous disc. The small disc, 
if the aperture of the mirror has a circular shape, takes the 
name of Airy disc. It is \textit{not} the geometrical image of the 
object. According to geometrical optics as the object integral 
with its lamp withdraws along the optical axis the concave mirror 
forms an image of it ever smaller and nearer the focal point. 
But the analysis applies to objects quite far from the horizon 
and, for essentially mathematical reasons, certainly doesn't 
extend to the case where the images are formed just in the focus. 
Indeed, geometrically the focus is a \textit{true point}. First of 
all it is not obvious which coordinates to assign to the object 
which is projected in the focus. Then although in the present 
case we know that the object on the horizon is extended we must 
believe that all its points admit of a single point as image 
and not only the one intersecting the optical axis. It is a question 
of definition if we consider the focal point an image of something 
or not. If yes, the anti-image must be posed by definition. Making 
the Abbe-Klein hypothesis the focus is included (with its antitransform). 
But Einstein claimed that to do physics it is necessary to measure 
with the double decimeter and with the clock with escapement 
and balance wheel. It is clear that the distance of the horizon 
cannot be measured even in a Gedankenexperiment by affixing a 
graduated line. With the words of Klein added at the foot of 
no. XXX of Vol. I for the 1921 edition, ``Occasionally Einstein 
opposed me with the following motivation: the transformation 
by reciprocal radii preserves the form of Maxwell's equations 
but not the relationship between coordinates and values measured 
with rulers and clocks''.

As mentioned, with the Abbe-Klein hypothesis the focus is included 
\cite{klein1968}. Mathematically it can be defined, 
``the anti-image of the focal point is a point''. Since the transformation 
is invertible, the anti-image also belongs to space, which topologically 
is simply connected.

Inclusion of the focus, \textit{just like its exclusion}, can be justified 
physically. We observed that Abbe found himself having to interpret 
the images especially when comparison with the object is not 
possible. In that predicament he imposes that the linear anti-image 
of the image contains all and only information accessible on the 
object. Clearly when the geometric image consists only of the 
focal point, even if we suspect or know that the object is not 
dot-like, from the received signal we can infer only that it 
is located along the direction of the optical axis in an indefinite 
position.

The interpretation given can be completed both physically and 
geometrically according to Abbe-Klein. The line of reasoning 
is the following. As the object withdraws from the mirror the 
light which it diffuses in that direction illuminates a broader 
zone. Thus the edge of the mirror diaphragms it under an ever 
narrower solid angle. The light \textit{apparently} emitted by the 
object is ever more collimated laterally or even more coherent 
spatially. If the object is sufficiently far the only light effectively 
received of all that diffused is the zero order of diffraction. 
It can be said that the aperture is too small to form an image. 
In this manner the reflected image, in addition to being, in 
agreement with geometrical optics, always smaller and nearer 
the focal point, is also always less detailed because it diffuses 
the only light the mirror receives.

We can assume that the mirror reveals on the principle of the 
holographic plate. Temporarily the analogy would appear with 
greater clarity if there were a way to receive a reference radiation 
also. Let us explain this. If the light coming from afar were 
sufficiently intense, let us say like sunshine, and if we put 
ourselves in the observation conditions of Newton, we would see 
rings due to the presence of the reference 
light. In the case of the mirror, a flat surface is lacking, 
pressed against the convexity of the mirror and capable of reflecting 
partially. Consequently the appearance of the image is rather 
uniform and even rather extended. It doesn't cover all of the 
plane because when a mirror is illuminated it also produces an 
image of itself, the exit pupil, to which geometrical optics 
can be applied in the same manner. Even if in the focal plane 
there can be neither more nor less light than that traversing 
the exit pupil, due to the place where the Airy disc is formed 
it is not its geometrical image in the conventional sense. Rather 
it has to do with a modulation of direction of the light originated 
by the mirror and superimposed on the light coming from afar. 
Fraunhofer's observed diffraction figure is mathematically the 
transfer function of the mirror, considered as a detector, in 
the sense that is attributed to the term in electrical engineering. 
We can agree that the diffraction image of the point alone would 
uniformly illuminate a plane (Abbe calls the relative condition 
``aplanatism'').

The diffraction pattern of the point, thought of as a geometrical 
plane, contains the same information as the focal image, thought 
of as dot-like; even in interferential terms the object is found 
along the direction of the optical axis, direction which in the 
absence of a reference radiation is arbitrary. In a word, we 
can say that a spatially coherent light in the absence of other 
light sources illuminates a very extended zone uniformly or that 
the image of the point on the horizon is ``delocalized'' to the 
fullest. Geometrically the previous definition is completed thus, 
``The transform of the anti-image is a plane''. Clearly it is 
not a plane of points. The geometry compatible with the double 
possibility of representing the focus point of a spherical mirror 
is the projective one. But the projective plane is a one-sided 
surface. Being it nonorientable, the mathematical distinction 
between left and right, as well as between up and down is without 
a difference.

\subsection{Experimental interpretation of the uncertainty}

We have described both the ``geometrical'' and the ``diffraction'' 
appearance of a point on the horizon. Given a spherical mirror 
and nothing else it does not correspond to two different observative 
possibilities. On the contrary, despite the interpretation being 
double a single image is observed, in each case an Airy disc. 
This disc limits by diffraction the quality of the images produced. 
In the case of the mirror, if the image belongs to a physical 
point, let us say a star, the disc presents itself accompanied 
by an edging in the form of a small number of more or less pale 
concentric rings which recall vaguely the photograph of a wave 
caused by the fall of a small object into a basin full of water. 
Detection of the direction light is coming from with respect 
to the principal optical axis of the mirror cannot be arbitrarily 
precise because of the modulation introduced by the finite aperture 
of the mirror. The spatial extension of the disc around the focal 
point can also be made plausible with the impossibility of exactly 
localizing a point on the horizon.

But the disc is also an effect of the aperture stop toward the 
light \textit{diffused by every point of a nearby object}. In this 
form it confers a more or less granular appearance on the images 
\cite{born1969}.

It is clear that the edged or granular appearance of the images 
has no counterpart in any basic structure of the object and/or 
light, which is revealed thanks to the choice of an experimental 
arrangement. Rather we can say that some uncertainty either about 
the direction of the light or about localization of a point in 
the spot which represents its image troubles this kind of optical 
measurement.

\subsection{Physical characterization of the objects requiring the 
point/plane geometrical duality}

In para. 8.4 we sought to make it physically plausible that 
two images correspond geometrically to a dot-like object. In 
geometrical approximation the two elements point and plane represent 
the above mentioned images. We introduced the `` object point'' 
as the anti-image of the image of a material object extended 
and illuminated with diffused light but on the horizon. We said 
that under these conditions the object is such that the mirror 
cannot magnify it. Since a plane is extended, we must clarify 
what is intended by ``magnification''. With this word Abbe designates 
the \textit{increase in discernable structural detail}. This means 
that the solid angle under which we see the object is in relationship 
with the information on the detail. Neither a structureless ball 
nor a homogeneous plate would ever be magnified according to Abbe. Vice versa, 
if relative position/direction are the only accessible information, 
any object behaves optically like a ball.

Perhaps it is worthwhile clarifying how the microscope lens functions 
according to Abbe using an analogy. Let us consider a radio signal. 
Everybody knows that the useful information is contained in the 
modulation while the carrier serves to allow tuning on the channel. 
Abbe presents the function of lens aperture exactly in these 
terms. He writes that the fineness of detail in the image des 
not depend on the luminousness of the collimated beam but on 
the modest percentage of light diffracted by the sample (\"{u}ber 
die Grenzen der geometrischen Optik. p.173). It is the magnification 
so understood that requires large lens numerical apertures and 
possibly homogeneous immersion lenses, typically in cedar oil. 
Returning to us, it is possible that Abbe didn't believe an unenlargeable 
object microscopic. But it seems to us that we can identify the 
unenlargeable object with the one with which de Broglie associates 
a dual nature.

\section{De Broglie's hypothesis in the context of electromagnetism}

If the focal point and its antitransform belong to the domain 
of transformation, to the dual anti-image-image transformation 
corresponds a shifting of the dot-like object from where it is 
to the focus of the mirror. We shall see below when the shifting 
indicated is a movement.

\begin{figure}
  \resizebox{8cm}{6.4cm}{\includegraphics{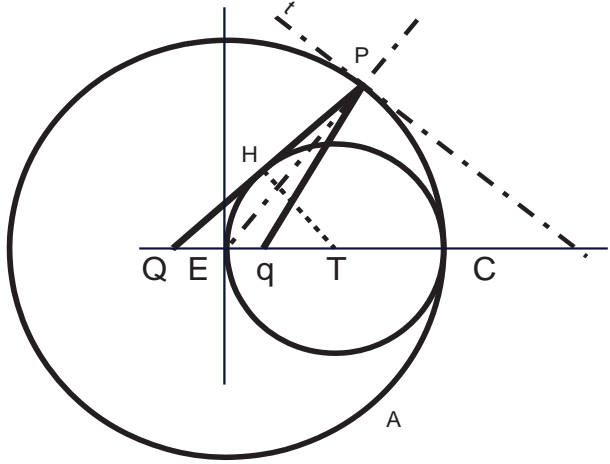}}
  \caption{Geometrical construction of the real image for spherical mirrors according to Newton.
ACP is the cross-section of the mirror with center E and vertex C; QC is the principal axis: t is the tangent at P.
Q is the object-point , q the image-point.
The angle of incidence of the ray QP, EPQ,
 and the reflection one, qPE satisfy the reflection law.
Despite the specific ray QP not being paraxial Newton's law still holds.
    }
  \label{fig_2}
\end{figure}

Let us assume first of all that the object is continuing to withdraw 
from us in accordance with any dynamic law of motion. According 
to Newtonian dynamics the withdrawing of the object can be unlimited 
and in this case a trajectory of not finite length is associated 
with it. But if we consider the path of the point representative 
of the image, it tends towards a point on the focal plane as 
limit point. The interpretation that the only optical information 
we are allowing for is on the relative direction of the movement 
of the object corresponds to this behavior of the image. But, 
due to the invertibility of the optical path, the path of the 
image is such for the object also. In other words, the geometrical 
support is identical for the path of the image and for that of 
the object. Let us assume now that reciprocally the object travels from the 
center of the mirror through the main focus point towards the 
vertex. This is a movement so that the support cannot have discontinuities 
in a neighborhood of the focal point. In fact the propagation 
trough the focus is not opposed by anything. As we noted above, 
when the mirror is spherical the representative points of the 
object and image on the path coincide in the center and vertex 
of the mirror. Then geometrically the unlimited withdrawing of 
the image must take place along a finite path and the common 
support of the two trajectories is homeomorphic with a circle.Experimentally it is very elusive to describe what happens to 
the image when the object illuminated in diffuse light passes 
through the focal point because in this zone the mirror amplifies 
to the maximum, the manufacturing imperfections are much enlarged 
and the increase in detail surely does not correspond to the 
magnification. This difficulty arises because a ``flux'' of the 
image through the focus is imagined. Not even when we can consider 
the image similar to the original is it admissible to call the 
light flux laminar. So if the focal point has not been whitened 
by the representation, hydrodynamics is difficult to apply. In 
addition, if the dynamic laws of motion could be applied to the 
image also it would not be reasonable to assume that in an arbitrarily 
short time interval, as measured by a clock, this might travel from where it is behind 
us to behind the reflecting surface. Thus in \textit{mechanics,} boundary 
conditions are specified on the reflecting surface, it is 
admitted that the image jumps and the specular reflection is 
not considered a movement. \textit{But the specular reflection is 
the effect which justifies recourse to non-mechanical formalisms}. 
Let us recall that the trajectory is not a fact but the result 
of a measurement operation. This measurement can be made by putting 
the ruler through each pair of positions successively occupied 
by the representative point and taking transit time into account, 
but can also be done by a theodolite. Remote ranging is an optical 
reception as much as specular reflection, only that it is applied 
to objects further away from the measuring instrument than the 
focal distance. If mathematical formalism can differ from that adopted in mechanics, 
the horizon aligned with the optical axis can be represented 
on the geometrical support as a transformed geometrical locus 
of the focal point and indicated by the symbol $\infty$. 
With this convention the marker point of the image no longer 
jumps but passes through that point. Since it is apparently not 
possible to find on the curve of support a neighborhood of  $\infty$ 
such that all the points are at a lesser distance therefrom than 
a predetermined length the trajectory is not measurable by means of a ruler along 
the entire path. Our present interpretation justifies Klein's 
warning (recalled in para. 7.1) that projective geometry would 
give rise to a theory of light completely different from Hamilton's.

The formalism foreign to the mechanical laws of motion exists 
already. In fact Klein shows by substitution of the analytical 
expression of the reflection transformation in Maxwell's equations 
that movements of this kind are solutions of it. The analytical 
form of the transformation is termed inversion or transformation 
by reciprocal radii. In Chap.XI, Vol.1 of his Treatise Maxwell 
himself deals with the same kind of solutions. In particular, 
being the theory of light dual, if the movement terminates in 
the focus, the solution also allows for de Broglie's hypothesis. 
As concerns material bodies the latter hypothesis says: ``It 
is impossible to ascertain the mechanical structure of an afar 
object by optical means just because of the assumed duality of 
the ethereal images''.

\subsection{Geometrical interpretation of Isaac Newton's formula for 
spherical mirrors}

We mentioned above that there has always been attributed 
to the constructions of which geometrical optics makes use the 
meaning of graphic prescriptions about a physical space. These represent 
the content of the formulas and describe the facts. But use of 
the constructions in mathematics is limited to exemplification 
in demonstrations and the solution of problems. There is no guarantee 
that a particular problem is not indeterminate or does not have 
infinite solutions. On the contrary, if a geometry represents 
a model of certain facts, the mathematical expressions of laws 
can rest directly on the model without the backing of more tried 
theories such as mechanics. In relation to the purely mathematical approach mentioned which 
was adopted at the end for light phenomena both by Einstein and 
Maxwell we show below that Newton's Axiom VI cas.2. represents 
a hyperbolic projective involution. In the cited axiom Newton transfers to a graph and calculates 
where a luminous axial point Q is reflected by a spherical mirror 
with center E and vertex C in the case of paraxial beams. Obviously 
since the light striking the mirror cannot be collimated using 
a diaphragm either the source Q is directional or the beams are 
deprived of any physical meaning. His prescription is: i) bisect 
any radius of the sphere, (suppose EC) in T; ii) let the point 
Q be the focus of the incident rays, the point q shall be the 
focus of the reflected ones; iii) you take the points Q and q 
so that TQ, TE and Tq be continual proportionals.

With his conventions,

\[
TQ:TE = TE:TqTq \ne 0
\]

The fixed point C in the figure \ref{fig_2} is practically 
coincident with the vertex of the mirror only for paraxial beams.

This formula expresses a Euclidean result. With reference to 
figure 2 the triangle QTH is rectangular at H by construction. 
Indeed, QH is tangent to the small circle of radius TE centered 
at T while TH is another radius of the same circle. But the tangents 
to the circumferences are orthogonal to the rays at the tangential 
point. The theorem of Euclid in question says that the square
$
{\mathop {{\rm (TE)}}\nolimits^{\rm 2} } 
$
 on the cathetus TH has an area equivalent to the rectangle 
which has for sides the hypotenuse QT and projection qT of the 
cathetus TH on the hypotenuse. When the formula is applied to 
mirrors both the distance of the object point and that of the 
image point are evaluated by the main focus of the mirror. For 
finite distances with a simple passage the formula for the power 
of the circle is written as follows (Coxeter loc. cit. p. 78):

\[
TQ \times Tq = \left( {TE} \right)^2 
\]

To find therefrom the transformation 
for reciprocal radii which Klein says it suffices to lay down,

\[
\left( {TE} \right)^2 = 1
\]

\begin{figure}
  \resizebox{8cm}{6cm}{\includegraphics{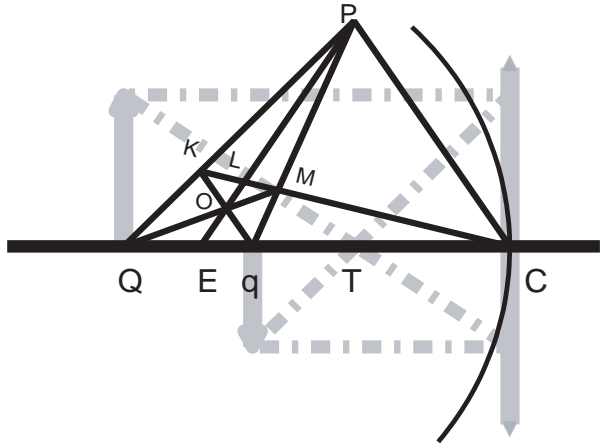}}
  \caption{Object-point Q and image-point q as conjugate harmonic points.
A quadrilateral PkQEqCMLO is shown in black on the most common graphic construction of the image.
The harmonic ratio among Q, q and the united points E and C does not depend on the choice of point P, provided it lays off the line CqEQ.
    }
  \label{fig_3}
\end{figure}

More recently by convention 
all the distances are referred to the vertex of the mirror which 
for paraxial rays is the point indicated here by C. E. Hecht 
shows in his book on optics what is the relationship between 
the two formulas and also specifies all the present conventions 
on signs. We continue to follow Newton's convention. In figure \ref{fig_3}
we show the relationship between the graphic 
construction of the image by means of light rays and the projective 
definition of the harmonic points on a straight line. The equation 
of mirrors relative to the figure is written,

\[
\frac{1}{{Cq}} + \frac{1}{{CQ}} = \frac{1}{{CT}} = \frac{2}{{CE}}
\]

This, rearranged as,

\[
\frac{{CE}}{{Cq}} - 1 = 1 - \frac{{CE}}{{CQ}}
\]

is the continuous harmonic proportional,

\[
\frac{{CE - Cq}}{{Cq}} = \frac{{CE - CQ}}{{CQ}}
\]

from which,

\[
\frac{{qE}}{{Cq}} = \frac{{EQ}}{{CQ}}
\]

\begin{figure}
  \resizebox{6cm}{7.2cm}{\includegraphics{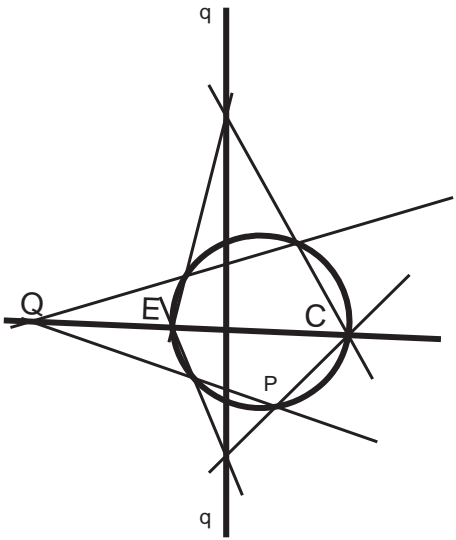}}
  \caption{The simplest projective correlation between a point and its plane.
A hyperbolic polarity is shown in order to compare it with Newton's axiom.
Q is the marker point of the object; qq is the section of its dual image; CPE is the section of the quadric (as a locus of points) that determines the polarity.
    }
  \label{fig_4}
\end{figure}

\begin{figure}
  \resizebox{8cm}{7.2cm}{\includegraphics{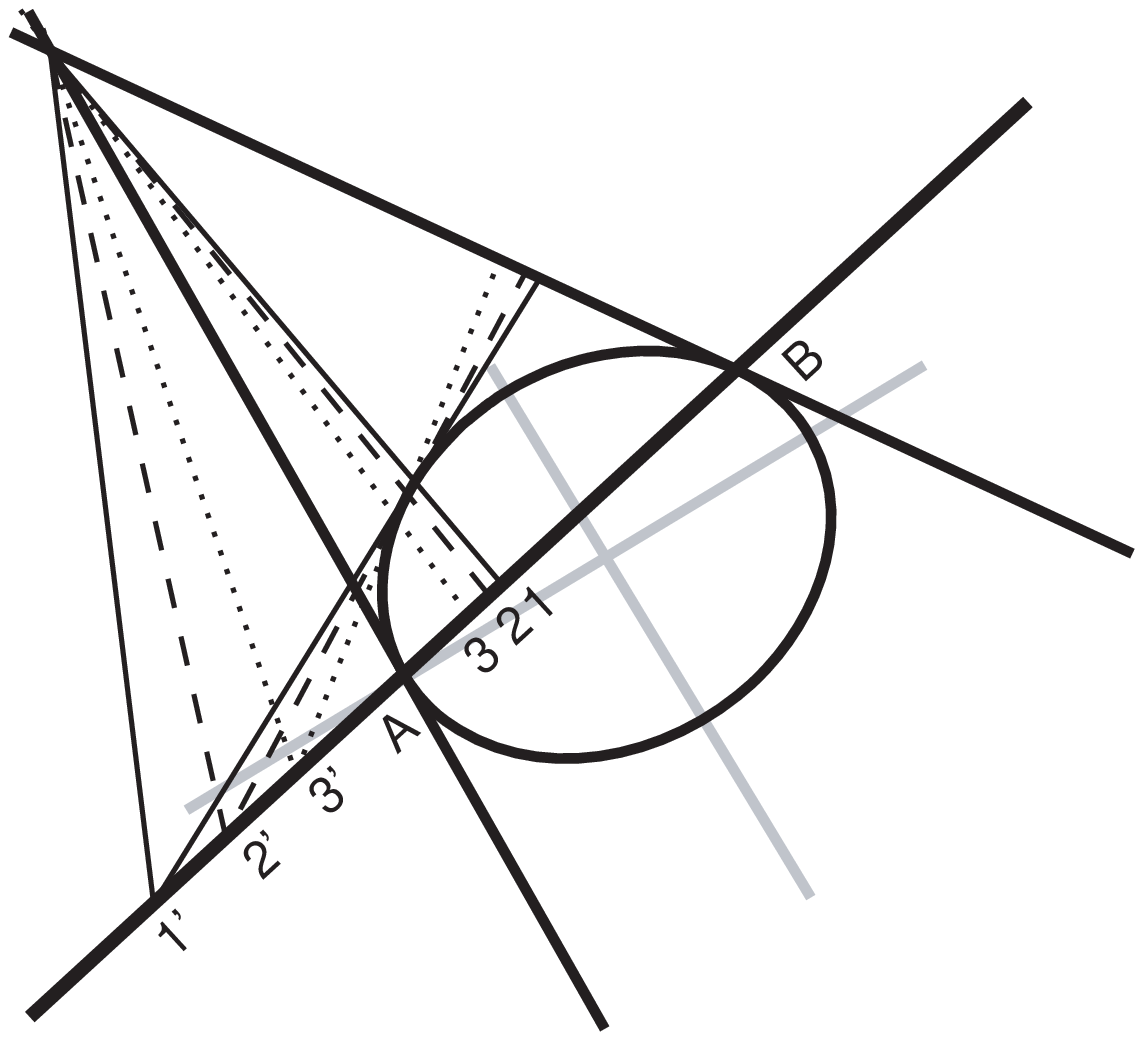}}
  \caption{The involution pertaining to spherical mirrors.
The relation establishes a double ordering involving each pair of points of a range.
1', 2', 3' ... locate the object Q; 1, 2, 3 ... locate the image q. A , B are invariant points.
The ordering might be induced from the quadric, as in stereographic projection.
    }
  \label{fig_5}
\end{figure}

With this cross ratio is associated 
the figure of the projective geometry which we have superimposed 
here on the usual graphic construction. If Q tends to the infinite, 
q tends to the mean point of EC, i.e. T, for a theorem relative 
to the mean point, without it being necessary to consider dynamic 
properties for light. The theorem is enunciated but in different 
geometries, depending on whether Q takes on the limit value or 
not. In this manner optics leans on one or the other geometry 
depending on how it interprets the phenomenology. If it interprets 
the diffraction image of Fraunhofer in the same manner as any 
other optical image, as we propose, going back to de Broglie, 
then the construction drawn in black causes a projective transformation.

While the full system of projective transformations either with 
the invariant quadrics (system of Reye) or assuming imaginary 
geometrical elements according to Klein is difficult to deal 
with geometrically, it is easy to show what the graphic appearance 
of the transformation involved is: polarity. We exemplify the 
construction in figure \ref{fig_4}. Finally in figure \ref{fig_5}
 we show the transformation implied in 
the case of spherical mirrors according to Newton. It associates 
a double ordering with the points of a range while leaving two 
points united.

\section{Summary and Conclusions}

In this work we have sought to specify a possible physical content 
for de Broglie's hypothesis. \textit{A priori} it could be associated 
with three distinct types of experimental observation, to wit,

i. comparison between spectral analysis of light and structural 
analysis of matter.
This aspect is the one thanks to which Bragg's experiments can 
be considered corresponding to those of Davisson and Germer. 
Despite this comparison having been most directly suggested by 
the hypothesis, \textit{a posteriori} the physical phenomena involved 
are too intricate to lend themselves to explain it physically.

ii. Dualism of reception, wavy or granular signal \cite{born1969}.
Depending on whether the electrical/optical detector used in 
the experiment was interferometrically locked to the signal or 
was rather a counter, the signal detected is interpreted as undulatory 
or corpuscular. In the classical schematic example of Young's 
device, if both slits are open and their image is projected correctly 
on a screen, delocalized interference fringes appear which the 
photomultiplier detects by counting. But if a single slit is 
open, the intensity modulation nevertheless observable on the 
screen and which gives rise to Fraunhofer diffraction does not 
determine the phase shift from the reference because the radiation 
which acted as reference was obscured. The photomultiplier also 
detects diffraction by counting.

iii. Dual solution.
According to us de Broglie himself proposed a conventional duality. 
But the mathematics he associated with it lacked whatever connection 
to experiment and was not coherent. By facing the problem in optics, where duality 
was first observed, one sees that the dual aspect of light lies 
in the fact, shown experimentally by Abbe, that image and diffraction 
pattern of microscopic objects have the same informative content. 
Abbe found this result by examining separately the contributions 
of the lens and of the eyepiece of the compound microscope. If 
this fact is expressed in terms of a coherent light-signal theory, 
then de Broglie's hypothesis is a hypothesis of mathematical 
nature formulated in the framework thereof. As a theory of light-signals 
we have considered Maxwell's. Instead of illustrating this theory 
with a mechanical analogy, we have leaned on geometry; we based 
on the Erlangen program to define movements. We sought to illustrate 
to what the basic point and plane elements of the model correspond 
and how they are transformed by specular reflection.

In the new context, de Broglie's hypothesis can be reformulated 
as, ``The same duality attributed to their images is attributed 
to material objects (i.e. to the antitransforms)''. This means 
that the theory of light, as far as the hypothesis applies, could 
be applied to movements of the material objects themselves.


\begin{thebibliography}{9}


\bibitem{messiah1999} A. Messiah, Quantum Mechanics, Dover Publications, Inc. (1999) New York. Orig. Ed. 1958

\bibitem{debroglie} L. deBroglie, Ann. de Phys. 10.s\'{e}rie,t.III p.22-128 (1925)

\bibitem{vanderwaerden} B. L. Van der Waerden, Sources of Quantum Mechanics, Dover Publications, Inc. (1968) New York

\bibitem{ditchburn1991} R. W. Ditchburn, Light, Dover Publications, Inc. (1991) New York. Orig. Ed. 1953

\bibitem{mehra2000} J. Mehra, H. Rechenberg, The Historical Development of Quantum Theory. Vol. 6 Part 1: The Completion of Quantum Mechanics, Springer Verlag (2000) New York

\bibitem{davisson1927} C. Davisson, L. H. Germer, Phys. Rev. 30(6):705-740 (1927)

\bibitem{einstein1917} A. Einstein, Phys. Z. 18:121-128 (1917)

\bibitem{einstein1907} A. Einstein, Ann. Phys 22:569-572 (1907)

\bibitem{towne1988} D. H. Towne, Wave Phenomena, Dover Publications, Inc. (1967) New York

\bibitem{einstein1905} A. Einstein, Ann. Phys 18:639-641 (1905)

\bibitem{debroglie1949} L. de Broglie in "Albert Einstein: philosopher-Scientist"  p. 107-127, Ed. P. A. Schilpp, Tudor Publishing Company (1949) New York

\bibitem{bohr1949} N. Bohr in "Quantum Theory and Measurement" p. 9-49, Ed. J. A. Wheeler, W. H. Zurek, Princeton University Press (1983) Princeton

\bibitem{feoli2002} A. Feoli, Europhys. Lett. 58(2):169-175 (2002)

\bibitem{abbe1989} E. Abbe, Abhandlungen \"{u}ber die Theorie des Mikroskops, Vol. 1 der Gesammelten Abhandlungen, Georg Olms Verlah (1989) Hildesheim

\bibitem{cushing2000} J. T. Cushing, Ann. Phys 9(11-12):939-959 (2000)

\bibitem{born1969} M. Born, Atomic Physics, 8. Ed. Dover Publications, Inc. (1969) New York Orig. Ed. 1935

\bibitem{matzkin2002} A. Matzkin, Eur. J. Phys. 23:285-294 (2002)

\bibitem{russell1981} B. Russell, I fondamenti della geometria, Club del libro Fratelli Melita (1981) La Spezia Orig. Ed. 1897

\bibitem{mach1982} E. Mach, Die Prinzipien der physikalischen Optik, historisch und erkenntnispsychologisch entwickelt, Minerva G.M.B.H. (1982) Frankfurt/Main Orig. Ed. 1913 (1921)

\bibitem{klein1973b} F. Klein, Gesammelte Mathematische Abhandlungen, Vol. 2, Nr. LXX p.601-602, Springer Verlag (1973) Berlin Orig. Ed. 1922

\bibitem{bailey2002} C. D. Bailey, Found. Phys. 32(1):159-176 (2002)

\bibitem{pauli1973} W. Pauli, Pauli Lectures on Physics, Vol. 2: Optics and the Theory of Electrons, Dover Publications, Inc. (2000) New York Orig. Ed. 1973

\bibitem{hasenoehrl1904} F. Hasen\"{o}hrl, Ann. Phys. 15:344-370 (1904)

\bibitem{manning1977} H. Manning Ed., The Fourth Dimension Simply Explained: A Collection of Essays Selected from Those Submitted in the Scientific American's Prize Competition (1909), Peter Smith (1977) Gloucester, Mass.

\bibitem{rucker1977} R. v. B. Rucker, Geometry, Relativity and the Fourth Dimension, Dover Publications, Inc. (1977) New York

\bibitem{maxwell1954} J. Clerk Maxwell, A Treatise on Electricity \& Magnetism, Vol. 1 Art. 53, 60 Dover Publications, Inc. (1954) New York Orig. Ed. 1891

\bibitem{klein1987} F. Klein, Vorlesungen \"{u}ber nicht-euklidische Geometrie, Verlag von Julius Springer 1928. Nachdruck 1968

\bibitem{boltzmann1999} L. Boltzmann, Modelli matematici, fisica e filosofia. Scritti divulgativi. - Sull'indispensabilit\'{a} dell'atomismo nella scienza (1897), Universale Bollati Boringhieri (1999) Torino

\bibitem{hund1984} F. Hund, Geschichte der Quantentheorie, 3. Auflage Wissenschaftsverlag, Bibliographisches Institut (1984) Mannheim

\bibitem{born1925} M. Born, P. Jordan, Z. Phys 34:858-888 (1925)

\bibitem{maxwell2001} J. Clerk Maxwell, Theory of Heat, Dover Publications, Inc. (2001) Orig. Ed. 1871

\bibitem{newton1979} I. Newton, Optics or A Treatise of the Reflections, Refractions, Inflections \& Colours of Light, IV Ed. Dover Publications, Inc. (1979) New York Orig. Ed. 1730

\bibitem{Maxwell1954b} J. Clerk Maxwell, A treatise on Electricity \& Magnetism, Vol.2 III Ed. Dover Publications, Inc. (1954) New York Orig. Ed. 1891

\bibitem{coxeter1989} H. S. M. Coxeter, Introduction to Geometry, II Ed. John Wiley \& Sons, Inc. (1989) New York Ed. 1961

\bibitem{reye1898} T. Reye, Die Geometrie der Lage, 1. Abteilung IV Ed. Baumg\"{a}rtens Buchhandlung (1898) Leipzig

\bibitem{klein1973a} F. Klein Gesammelte mathematische Abhandlungen, p. V \& p. 460 Springer- Verlag (1973) Berlin Orig. Ed. 1921

\bibitem{pohl1976} R. W. Pohl, Optik und Atomphysik, 13. Auflage Springer-Verlag (1976) Berlin

\bibitem{hecht1998} E. Hecht, Optics, III Ed. Addison-Wesley (1998) Reading, Mass.

\bibitem{klein1968} F. Klein, Elementarmathematik vom h\"{o}heren Standpunkte aus, Vol. II: Reliefperspektive, p. 102, Verlag von Julius Springer (1925) Berlin Nachdruck 1968





\end{thebibliography}
\end{document}